\documentclass{solarphysics}
\usepackage[optionalrh,solaromanenum]{spr-sola-addons} 
\usepackage{graphicx}                    
\usepackage{color}                       
\usepackage{url}                         

\newcommand{\aap}{    {\it Astron. Astrophys.}}

\newcommand{\apj}{    {\it Astrophys. J.}}

\newcommand{\apss}{   {\it Astrophys. Space Sci.}}

\newcommand{\solphys}{{\it Solar Phys.}}

\begin{document}

\begin{article}

\begin{opening}

\title{Fine and superfine structure of Decameter-Hectometer type II burst on 2011 June 7}

%
\author{V.~V.~\surname{Dorovskyy}$^{1}$\sep
        V.~N.~\surname{Melnik}$^{1}$\sep
        A.~A.~\surname{Konovalenko}$^{1}$\sep
        A.~I.~\surname{Brazhenko}$^{2}$\sep
        M.~\surname{Panchenko}$^{3}$\sep
        S.~\surname{Poedts}$^{4}$\sep
        V.~A.~\surname{Mykhaylov}$^{5}$}

%

\institute{$^{1}$ Institute of Radio Astronomy
                     email: \url{dorovsky@rian.kharkov.ua}\\
              $^{2}$ Poltava Gravimetric Observatory
                     email: \url{brazhai@gmail.com}
              $^{3}$ Space Research Institute
                     email: \url{mykhaylo.panchenko@oeaw.ac.at} \\
              $^{4}$ University KU Leuven
                     email: \url{Stefaan.Poedts@wis.kuleuven.be} \\
              $^{5}$ Karazin National University  }

\begin{abstract}

The characteristics of the type II bursts with herringbone
structure observed both by ground based radio telescopes (UTR-2,
URAN-2) and spaceborn spectrometers (STEREO A-B) are discussed.
The burst was recorded on 7 June, 2011 in the frequency band
3--33~MHz. It was characterized by extremely rich fine structure.
The statistical analysis of more than 300 herringbone sub-bursts
constituting the burst was performed separately for the positively
(reverse) and negatively (forward) drifting sub-bursts. The sense
and the degree of circular polarization of the herringbone
sub-bursts were measured in the respectively wide frequency band
(16--32~MHz). A second order fine frequency structure of the
herringbone sub-bursts was firstly observed and processed. Using
STEREO COR1 (A,B) and SOHO LASCO C2 images the direction and
radial speed of the CME responsible for the studied type II burst
were determined. The possible location of the type II burst source
on the flank of the shock was found.
\end{abstract}

%

\end{opening}

%
\section{Introduction}
    \label{s:Introduction}

The beginning of the 24-th solar cycle was accompanied by the
increased rate of high energy transients, such as Coronal Mass
Ejections (CME). It looks natural since CMEs are considered to be
tightly connected with the solar flares \cite{Yashiro08} and the
flare occurrence rate is the signature of the solar activity. CMEs
belong to the class of events which play the key role in the space
weather forming. One of the CME signatures in radio band are type
II bursts. It is currently accepted that type II bursts are
generated by sub-relativistic electrons accelerated at the front
of the shocks driven by CMEs \cite{Zaitsev98,Mann05,Miteva07}.
Type II bursts are observed in the wide frequency range from
decimeter (dm) through meter (m) and decameter-hectometer (DH)
down to kilometer (km) wavelengths bands. In frames of the plasma
emission mechanism this fact proves the existence of shocks at the
heliocentric distances from one solar radius up to 1 AU. At high
frequencies (e.g. dm and m bands) type II burst may also be
connected with the flare-generated blast waves without associated
CMEs \cite{Jasmina12}. But observations show that these
flare-generated shocks are characterized by fast deceleration and
dissipation \cite{Vrsnak08} and hence cannot propagate high in the
corona. At the same time type II bursts which can be traced
consequently in m, DH bands and lower are connected with shocks
driven by CMEs of extremely high kinetic energy thus being
potentially the most geoeffective events \cite{Gopalswamy05}.

From this point of view and taking into account the possibility of
ground based observations with excellent sensitivity and
resolutions, the detailed study of type II bursts parameters in
the decameter wavelengths band is of great scientific interest.

Special attention should be focused on study of type II bursts
with  so called "herringbone" (or simply "HB") structure since
they give the most obvious manifestation of the electrons
acceleration at the shock front \cite{Zaitsev98}.

The HB structure was discovered and described by \cite{Roberts59}
in meter wavelengths band. The term "herringbone" comes from the
morphology of the dynamic spectrum of the burst, representing the
positively and negatively drifting short sub-bursts diverging from
the main backbone towards higher and lower frequencies
respectively and reminding the fish-bone. The type II bursts with
HB structure are not rare events. Cane and White
(\citeyear{Cane89}) stated that in average 21$\%$ of all type II
population exhibit HB structure. They also noted that the HB
structure occurrence rate is strongly correlated with the
intensity. They also remarked that in a group of intense type IIs
as many as 60$\%$ of them have HB structure.  HB structure as well
as normal type II bursts exhibit fundamental and harmonic
emission. Fundamental HB sub-bursts are more intense and more
strongly polarized than the fundamental ones, unlike the maternal
type II burst, whose fundamental emission is weaker and both
harmonics are almost unpolarized \cite{Cairns87}.
\citeauthor{Suzuki80} (\citeyear{Suzuki80}) noted also that
negatively drifting HB sub-bursts tended to have higher degree of
circular polarization than the positively drifting ones.

\section{Observations and analysis}
    \label{s:Observations}

In summer months 2011 the observations of the solar sporadic radio
emissions  were performed by two Ukrainian decameter radio
telescopes: the UTR-2 radio telescope (Southern arm with
collecting area $\approx$ 50000~m$^2$) \cite{Braude78} and the
URAN-2 array with total collecting area of $\approx$ 28000~m$^2$
\cite{Brazhenko05}. Since the URAN-2 telescope consisted of
cross-dipoles it allowed measuring the sense and the degree of
circular polarization of the radio emission. Both radio telescopes
operated in frequency range 8 -- 32~MHz and were equipped with
similar up-to-date back-ends. Dual channel spectropolarimeter DSPZ
performed real time 8192-point FFT each 120 $\mu$s and provided
real time polarization degree calculations. The two telescopes
were separated by $\approx$ 150~km that allowed to exclude the
effect of the ionosphere.

For the analysis we have selected the type II bursts observed on 7
June 2011. This burst was registered by both the ground based and
spaceborn radio telescopes in frequency band 3--32~MHz. The type
II burst itself was a part of more complex and large-scale event
which apparently was initiated by the powerful solar M2 flare
occurred at 06:20~UT near NOAA11226 active region (S22W52).

\begin{figure}
\centerline{\includegraphics[width=0.9\textwidth,clip=]{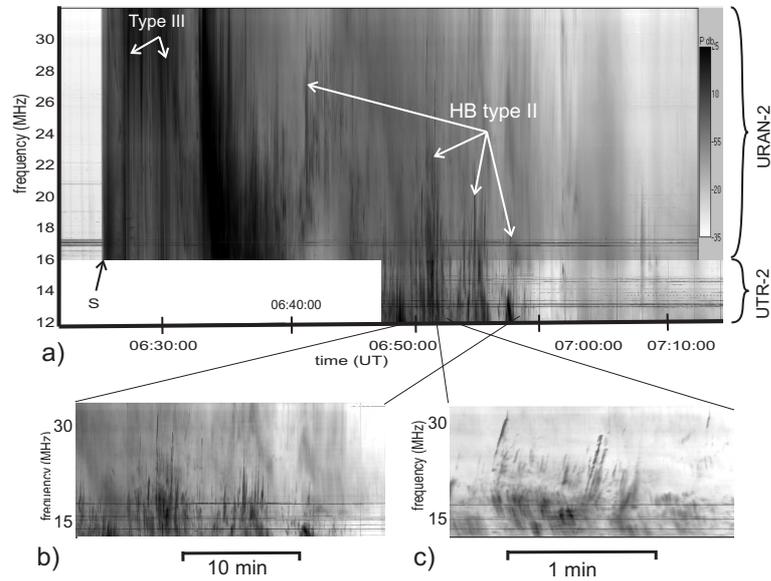}}
\caption{Dynamic spectra of: the whole event (a), the wavelike
backbone (b) and separate HB sub-bursts (c). } \label{fig:1}
\end{figure}

The dynamic spectrum of the whole complex event observed by URAN-2
and UTR-2 radio telescopes on 7 June 2011 is shown in Figure
\ref{fig:1}a. The event started at 06:29~UT with a sudden increase
of the radio emission continuum by almost 5 orders of the flux
magnitude in the whole frequency range covered by the radio
telescopes.  The continuum jump is marked with $s$. About 90
minutes later a group of powerful type III bursts was registered.
Finally, the type II burst with peak fluxes reaching $10^6$~s.f.u.
started at 06:34:00~UT and lasted for about half an hour.

In addition we must note that this type II burst was briefly
described by \citeauthor{Zucca12} (\citeyear{Zucca12}). They made
observations in frequency band 20-400~MHz with Callisto
spectrograph installed at Rose Solar-Terrestrial Observatory and
found that the burst consisted of Fundamental and Harmonic
components and was accompanied by moving type IV continuum at
frequencies 130-400~MHz. Unfortunately insufficient time and
frequency resolution of the experiment did not alow detailed
analysis of the rich fine structure of the type II burst. On the
contrary UTR-2 and URAN-2 radio telescopes provided data with
frequency resolution of 4~kHz and time resolution of 100~ms in
frequency bands of 12 -- 32~MHz and completely resolve the
emission fine structure.

Zoomed fragment of the dynamic spectrum (Figure \ref{fig:1}c)
unambiguously shows that this burst belongs to the well known
class of type II bursts with HB structure since separate
sub-bursts of this structure with positive and negative frequency
drift rates are clearly seen there. Standard type II bursts of
such kind consist of separate ``bones'' with drift rates of the
opposite signs and the ``backbone'' -- the emission lane from
which the ``bones'' seem to start. However there were reports
about "backboneless" HB structured type IIs \cite{Holman83}, when
the backbone looks like an emission gap separating the sub-bursts
with drifts of the opposite signs. The latter was the case on 7
June 2011.

\subsection{Frequency drift rate of the type II backbone}
    \label{s:drift rate}

\begin{figure}
\centerline{\includegraphics[width=0.7\textwidth,clip=]{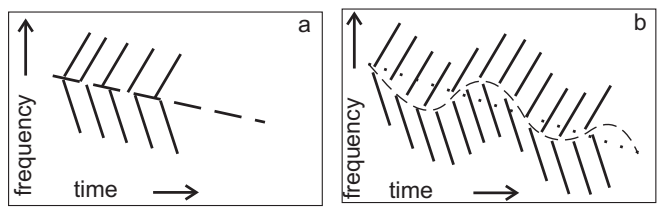}}
\caption{Schematic view of normal HB type II (a) and type II burst
with the wavelike backbone (b) } \label{fig:2}
\end{figure}

Usually the frequency drift rates of the backbone of type II
bursts with HB structure are either very low \cite{Melnik04} or
absent at all \cite{Carley13}. This fact is commonly explained by
the shock wave propagation in non-radial direction, i.e. at the
substantial angles ($>45^\circ$ ) to the coronal plasma density
gradient \cite{Holman83,Melnik04}.  The discussed type II burst is
out of the common. Its backbone has not only respectively high
average frequency drift rate but apparently "wavelike" appearance
in the dynamic spectrum as schematically shown in Figure
\ref{fig:2}b by dashed line.

In the case of monotonous backbone frequency drift (Figure
\ref{fig:2}a)  the drift rate is determined by the radial velocity
of the CME-driven shock. In the case of "wavy" backbone (Figure
\ref{fig:2}b) it is reasonable to estimate average radial velocity
of the source by the trend of frequency drift rate  along the line
representing the linear approximation of the backbone (dotted line
in Figure \ref{fig:2}b).

For the discussed type II burst the average drift rate is about
-25  kHz/s within the frequency bounds 16--25~MHz that corresponds
to radial component of the shock front velocity near
600~km~s$^{-1}$. Hereinafter all radial velocities are given for
Newkirk corona model. At the same time instant absolute drift
rates of separate parts of the backbone sometimes exceed
100~kHz~s$^{-1}$. Totally we have registered 3 full periods of the
backbone oscillations between 6:41~UT and 6:57~UT. These periods
are 530~s, 230~s and 150~s respectively. The magnitude of these
oscillations along frequency axis averaged 10~MHz.

\begin{figure}
\centerline{\includegraphics[width=0.9\textwidth,clip=]{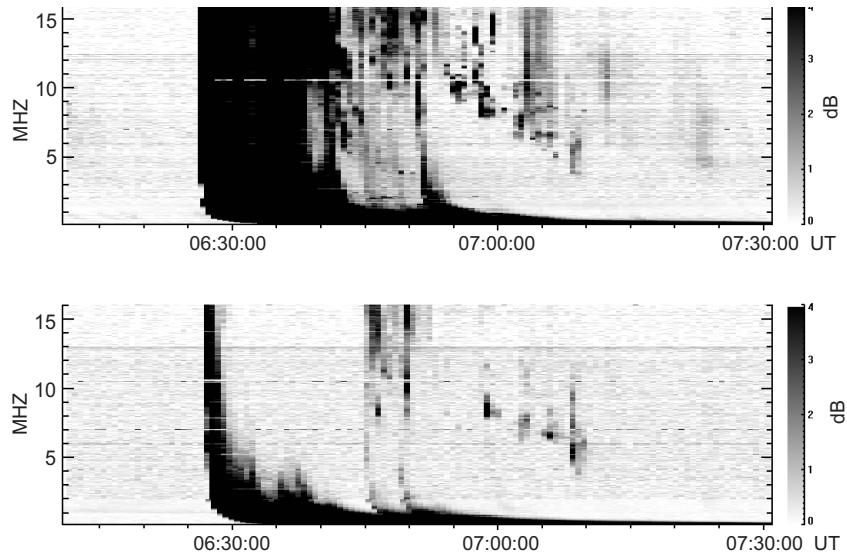}}
\caption{Type II burst from STEREO-A (top panel) and STEREO-B
spacecraft (bottom panel).} \label{fig:3}
\end{figure}

The continuation of the burst at frequencies below 12~MHz  was
registered by the STEREO/WAVES radio receivers. Identification of
the HB structure and thus detection of the backbone in data
obtained from the spaceborn receivers was not possible due to bad
time resolution (1~min) and insufficient sensitivity. So we
derived the total frequency drift rate along the low-frequency
edge of the dynamic spectrum (Figure \ref{fig:3}).

The obtained drift rates of -8~kHz~s$^{-1}$ at 6~MHz and -3
kHz~s$^{-1}$ at 3~MHz correspond to the linear velocities of 550
and 400~km~s$^{-1}$ respectively.

High frequency and time resolutions of the ground-based  back-ends
allowed to make detailed statistical analysis of the main
parameters of the HB structure sub-bursts. During the life-time of
the type-II bursts about 300 separate sub-bursts were identified,
200 of which had negative frequency drift rates (forward
sub-bursts) and the rest had positive ones (reverse sub-bursts).
    The basic parameters of the sub-bursts such as frequency drift  rates and durations appeared to be distributed within relatively narrow intervals.

\subsection{Frequency drift rates of the HB sub-bursts}
    \label{s:HB_drift}

\begin{figure}
\centerline{\includegraphics[width=0.9\textwidth,clip=]{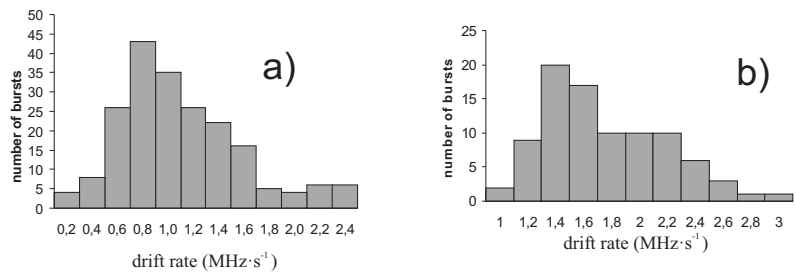}}
\caption{Distributions by the frequency drift rates of the forward
(a) and the reverse HB sub-bursts (b).} \label{fig:4}
\end{figure}

Total amount of the forward sub-bursts counted 201. The distribution of these sub-bursts by the absolute value of the drift rate is given in Figure~\ref{fig:4}a. Figure~\ref{fig:4}b shows corresponding distribution of the reverse sub-bursts.

The average absolute value of the forward sub-bursts drift  rates
appeared to be equal to 1.23~MHz~s$^{-1}$ with maximum of
distribution at 0.8~MHz~s$^{-1}$. The distribution is evidently
asymmetric with steep slope towards slower drift rates and rather
flat fall towards faster ones. This distribution resembles those
obtained for forward drift pairs \cite{Melnik05} and solar
S-bursts \cite{Dorovskyy06,Briand08}.

The corresponding distribution of the reverse sub-bursts in
general is close to that for forward ones, except the absolute
average drift rate which is 1.5 times higher (1.8~MHz~s$^{-1}$)
and maximum of the distribution observed at 1.4~MHz~s$^{-1}$. From
above figures we may conclude that the drift rates of the HB
sub-bursts of one separate type II burst are distributed in a
respectively narrow range from 0.5 to 2.5~MHz~s$^{-1}$, that is
slightly slower than drift rates of normal type III bursts in
decameter wavelengths band.

Since the HB structure was observed in the respectively wide
frequency band, it is of evident interest to investigate the
dependence of the absolute drift rates of the sub-bursts on
frequency.

The obtained dependences were approximated by the power low
empirical equations separately for forward (1) and reverse (2)
sub-bursts:

\begin{equation}  \label{Eq_1}
\left| \frac{df_f}{dt} \right|_{F}=0.01\cdot f^{1.73}, \quad \mbox{standard  error = 0.12},
\end{equation}

\begin{equation}  \label{Eq_2}
\left| \frac{df_r}{dt}\right|_{R}=0.088\cdot f, \qquad \mbox{standard error = 0.15}.
\end{equation}

Here drift rates are given in MHz~s$^{-1}$ and frequency in MHz.
The dependence for forward sub-bursts (\ref{Eq_1}) is very close
to that derived for type III bursts by \citeauthor{Alvarez73}
(\citeyear{Alvarez73}) while the dependence for reverse HB
sub-bursts (\ref{Eq_2}) appears to be linear.

\subsection{Durations of the HB sub-bursts}
    \label{s:HB_duration}

\begin{figure}
\centerline{\includegraphics[width=0.8\textwidth,clip=]{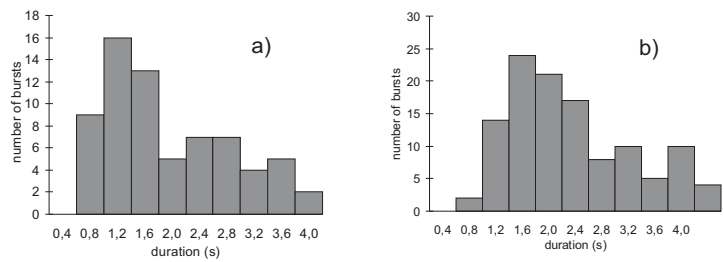}}
\caption{Distributions by the instant durations of the forward HB
sub-bursts (a) and the reverse HB sub-bursts (b).} \label{fig:5}
\end{figure}

Another important parameter of the solar radio bursts carrying
the information about the properties of the sub-relativistic
electron beams is burst duration at fixed frequency. The
distribution of the HB sub-bursts by their instant durations is
shown in Figure \ref{fig:5}. Roughly speaking these distributions
are similar. The only difference is the average durations of
forward (2.3 s) and reverse (1.9 s) sub-bursts. Both values are
substantially smaller than typical durations of the type III
bursts.

\subsection{Polarization of the HB sub-bursts}
    \label{s:HB_polarization}

\begin{figure}
\centerline{\includegraphics[width=0.7\textwidth,clip=]{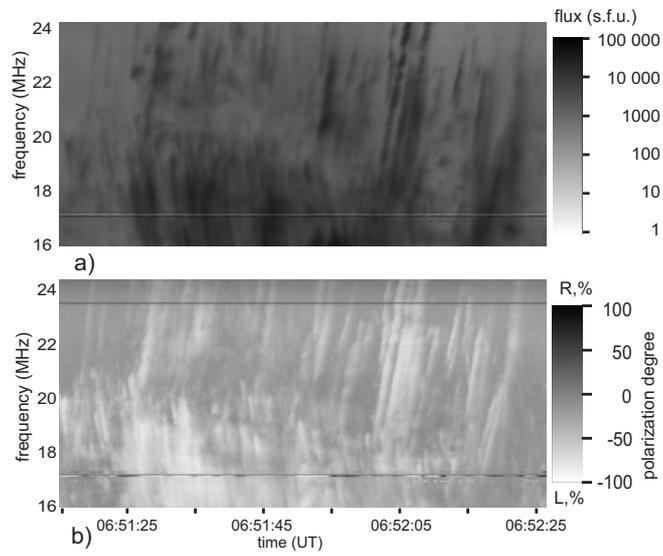}}
\caption{Fragments of the power (a) and the polarization spectra
of the HB structure (b).} \label{fig:6}
\end{figure}

The URAN-2 radio telescope equipped with the broad-band
spectropolarimeter  allowed us to detect the sense and the degree
of circular polarization of the sub-bursts in the frequency band
16--32~MHz. Investigations of the HB structure polarization was
firstly performed at such low frequencies. The power and
polarization spectra are shown in Figure \ref{fig:6}.

As was noted earlier \cite{Suzuki80} the HB structure sub-bursts
corresponding to the fundamental emission usually have
considerably higher degree of circular polarization in comparison
with the polarization degree of the backbone (if it exists) and
parent type II bursts. Our observations confirm this fact. Average
degree of circular polarization of both kinds of the sub-bursts
equaled 50$\%$ in some cases reaching as high as 80$\%$.  It seems
very important that the sense of circular polarization of the
forward and reverse sub-bursts appeared to be the same. In all
cases the left-handed circular polarization was observed. At the
same time the polarization of the preceding powerful type III
bursts was of much less degree (20--30$\%$) and of the opposite
sense.

\subsection{Fine frequency structure of the HB sub-bursts}
    \label{s:HB_fine_structure}

\begin{figure}
\centerline{\includegraphics[width=0.7\textwidth,clip=]{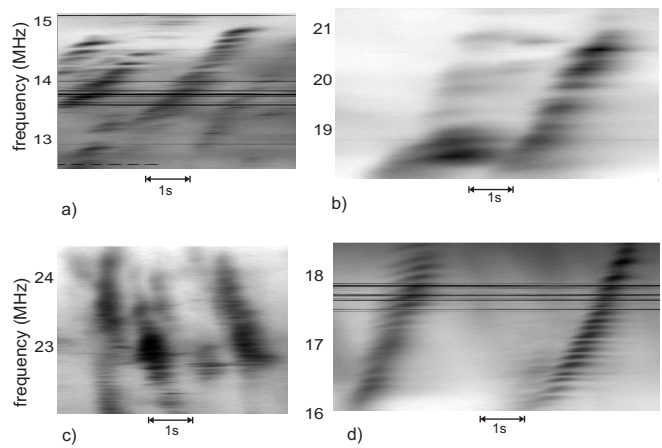}}
\caption{The fine frequency structure of the HB sub-bursts.}
\label{fig:7}
\end{figure}

High frequency resolution of the experiment allowed to define a
"second order fine structure" of the HB sub-bursts. This fine
structure (see Figure \ref{fig:7}) has an appearance of the
quasi-periodic (in frequency domain) chain of narrow-band
sub-bursts. The bandwidths of these sub-bursts were 30--60~kHz and
frequency spacing between neighboring sub-bursts were 60--120~kHz.
This fine structure is close to the "frindge" structure of solar
S-bursts, described in \cite{McConnell82}. As a rule the
narrow-band sub-bursts had no own drift rate while in rare cases
the slight positive drift rates of 100~kHz~s$^{-1}$ were observed
(Figure \ref{fig:7}d). The characteristic feature of the first (in
time) sub-burst shown in Figure \ref{fig:7}d is that the drift
rate of the narrow-band sub-bursts changes from +150~kHz~s$^{-1}$
at the beginning down to 20~kHz~s$^{-1}$ at the end of the burst.

\subsection{The associated solar events}
    \label{s:associated_events}

Data obtained from SOHO show that the discussed complex solar
burst was the result of the M2 solar flare, taken place near the
active region NOAA11226 located in the South-West part of the
solar disk (22$^{\circ}$S, 52$^{\circ}$W). According to the GOES
data the X-Ray flux was raising from the background level till the
maximum point for 8 min, from 06:19~UT till 06:27~UT. Along with
it the flux of the background continual solar radio emission was
rapidly increasing from quiet Sun level to the value of 10000
s.f.u. for only 9~s -- from 06:26:03 till 06:26:12~UT.  We must
note that the leading edge of this increase had characteristic
drift rate of -4.5~MHz~s$^{-1}$ from high to low frequencies. It
indicates that the discussed continuum raise could be caused by
the electrons moving  with velocity of
$\sim$10$^{10}$~cm~s~$^{-1}$ away from the Sun. The electrons with
such velocities are usually associated with normal type III
bursts.  Approximately in 2 minutes after the continuum raise the
dense group of powerful type III bursts started. Taking into
account the time needed for electrons to travel from the flare
region to the location of the observed radio emission source (at
the velocity of 10$^{10}$~cm~s$^{-1}$ it roughly counts 5~s), we
can conclude that acceleration of the electrons responsible for
the continuum jump took place during the raising phase of the
flare, before X-ray flux reached its maximum.  At the same time
the electrons which initiated the subsequent type III bursts were
accelerated after the flare reached the maximum intensity. We
should also note that such a sequence when type II burst is
preceded by the group of powerful type III bursts and followed by
type IV continuum is observed respectively often.

\begin{figure}
\centerline{\includegraphics[width=0.8\textwidth,clip=]{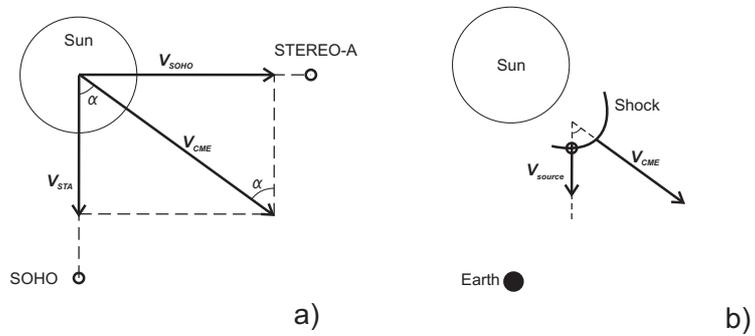}}
\caption{The geometry of CME speed detection (a) and schematic
view of the
type II source movement (b).} \label{fig:8}
\end{figure}

Using data of COR1 (A and B) coronagraph installed onboard the
STEREO satellites it is easy to define that the CME driven shock
reached the heliocentric height of 1.7R$_\odot$ with local plasma
frequency of 30~MHz at about 06:30~UT. This time corresponds to
the onset of the type II bursts at frequency 30~MHz. Since we've
got sky-plane images of the CME obtained from 3 different
coronagraphs located along the Earth's orbit at approximately
90$^{\circ}$ angular separation it was possible to retrieve the
real absolute velocity and the direction of the CME propagation as
shown in Figure \ref{fig:8}a. The sky-plane component of the CME
velocity obtained from STEREO-A spacecraft ($V_{_{STA}}$) was
found to be equal 1200~km~s$^{-1}$ and accordingly that from SOHO
spacecraft ($V_{_{SOHO}}$) equaled 1550~km~s$^{-1}$. Solving
simple trigonometric equations we obtain the absolute CME velocity
($V_{_{CME}}$) of 1950~km~s$^{-1}$ and the direction of about
52$^{\circ}$ westward from the observer's line-of-sight.
Apparently the obtained angle coincides with the longitudinal
position of the active region NOAA11226 and proves the previously
supposed association between the CME and the active region. This
fact also agrees with the statement that CMEs preferably propagate
in the radial direction above the corresponding active region
\cite{Yashiro08}.  At such a velocity the CME would cover 1 AU
distance in approximately 20 hours, and thus should arrive to the
Earth's orbit around 02:30 UT on 8 June 2011.

It is known that the ability of CME to produce considerable
impacts upon the space weather conditions is determined by its
basic kinetic properties: the speed, the angular width and the
direction \cite{Gopalswamy11}. The population of the geoeffective
CMEs has average speed of $\sim$1000~km~s$^{-1}$ and includes
mostly halo-CMEs which are in fact normal CMEs propagating towards
(or straight away) the Earth \cite{Gopalswamy09}. In addition many
authors reported \cite{Michalek06,Kim10} that at all other
parameters being equal the CMEs originating from western
hemisphere used to cause more intense geomagnetic storms.
Discussed halo-CME had all signs of geoeffectiveness except the
direction. Nevertheless slight decreasing of the Dst index down to
-30~nT in average was registered at the beginning of the 8 June
2011. We suppose that this decrease might be caused by the eastern
flank of the CME when it reached the Earth's magnetosphere

At the same time the solar wind parameters deviations were
observed by the ACE and GOES spacecraft at the time of the CME
possible arrival. Thus, the proton speed increased from 400 to 550
km~s$^{-1}$, proton temperature raised from 7$\cdot$10$^4$~K to
2$\cdot$10$^5$~K and shortly, for only 5 hours there was a jump in
the proton density from 4 to 12~cm$^{-3}$.

\section{Discussion}
    \label{s:Discussion}

As it follows from the data analysis the shock linear velocity
retrieved from coronagraph observations is at least three times
higher than the radial velocity of the radio source derived from
the frequency drift rate of the burst (1950 and 650~km~s$^{-1}$
respectively). In our opinion this discrepancy could take place
when the source of the type II bursts was located not at the
forehead but at the flank of the shock. Indeed, the drift rate of
the bursts is determined by the radial component of the total
linear velocity of the source, as

\begin{equation} \label{Eq_3}
D_f=\frac{f}{2}\frac{1}{N}\frac{dN}{dr}\cdot V_s \cdot \cos{\theta},
\end{equation}

where $f$ is the frequency, $N$ is the plasma density, $dN/dr$
isthe density gradient, $V_s$ is the source linear velocity and
$\theta$ is the angle between the density gradient and the source
velocity  vector. In other words only the radial component of the
total source velocity contributes to the frequency drift rate.
Moreover the flanks of CME where the emission source is supposed
to be located may propagate slower than the CME forehead
\cite{Gopalswamy09}. Detailed analysis of the SOHO C2 images shows
that the radial velocities of the regions, deflected from the
forehead direction by approximately 40--50$^{\circ}$ (in
latitudinal plane) are at least one third of the CME forehead
speed. Assuming quasi-symmetrical cone-like expansion of the CME
\cite{Gopalswamy09} one may suppose the same velocity for the case
of longitudinal deflection by the mentioned angle. Taking into
account the derived CME propagation direction we conclude that the
type II burst may originate from the eastern flank of the shock,
which moves non-radially ($\sim$50$^{\circ}$  to the density
gradient) towards the Earth as shown in Figure \ref{fig:8}b.

The wavelike oscillations of the backbone was firstly noted by
\cite{Melnik04}. Authors assumed that "waving" could be the result
of the shock front intersecting coronal streamer while moving
almost normally to the coronal density gradient. Recent
investigations, e.g. \cite{Reiner03,Feng12,Kong12}, show that the
sources of type II radio emissions are likely located at regions
where CME flanks intersect dense coronal streamers. This
explanation seems reasonable also for the discussed burst. With
little corrections that the source apparently moved at the angle
about 50$^{\circ}$ to the CME forehead direction and intersected
at least three inhomogeneous regions, most likely streamers. The
magnitude of the backbone oscillations in this case is determined
by the ratio between plasma density inside the streamer and the
ambient coronal plasma density. Thus 10 MHz oscillation magnitude
corresponds to the density deviation of 5$\cdot 10^6$~cm$^{-3}$.
Judging from the periods of the oscillations the linear distance
between neighboring streamers are 0.5~R$_{\odot}$,
0.2~R$_{\odot}$, and 0.15~R$_{\odot}$. Since these distances are
determined at the heliocentric heights of about 2~R$_{\odot}$ the
corresponding angular separations between the streamers count
$\sim 10^{\circ}, 15^{\circ}$ and $25^{\circ}$. Then transverse
dimensions of the streamers at the mentioned height vary from
$\sim$0.1 to 0.2  R$_{\odot}$.

High degree of circular polarization of the herringbone sub-bursts
indicates the fundamental emission. On the contrary the
polarization of the accompanying type III has opposite sense. It
may indicate that the conditions at the sources regions of the HB
and type III bursts are quite different. First of all it concerns
the magnetic field direction.

The frequency range over which a separate HB burst extends is
rather narrow in comparison with type III bursts. For the observed
burst this frequency range varied from 2 to 10~MHz. Taking into
account the source velocity derived from the HB bursts drift rates
linear dimensions of the area when emission is possible lay
between $3\cdot 10^9$~cm and $2\cdot 10^{10}$~cm (or
0.05~R$_{\odot}$ and 0.3~R$_{\odot}$).

It is commonly assumed the linear velocities of the electron
generating the forward and the reverse HB bursts are approximately
equal. The difference between mean values of the absolute drift
rates of these two populations are more likely caused by different
conditions behind and in front of the source position. This fact
may also play role in the difference of the drift rates
dependences (\ref{Eq_1}) and (\ref{Eq_2}). One can easily note
that the dependence for forward HBs is more close to the universal
dependence $D_f=-0.01\cdot f^{1.84}$ derived by Alvarez and
Haddock (\citeyear{Alvarez73}) for the type III bursts in the
undisturbed corona then that for reverse ones. It seems natural
since the forward HBs are produced by electrons moving away from
the Sun, across the undisturbed yet corona. The electrons
propagating towards the Sun meet the after-shock corona and thus
the density gradient may be different from that of quiet corona.

\section{Conclusion}
    \label{s:Conclusion}

The detailed analysis of the type II burst observed on 7  June
2011 by three different radio telescopes was performed.

This type II burst occurred in association with the preceding
powerful type III bursts, which seem to be initiated by the solar
flare near AR11226. The total frequency drift rate of the burst is
close to that of ordinary type II. Thus drift rates of HB type II
bursts may vary from several tens of kHz per second in the
decameter band down to zero.  Sometimes such bursts may have no
backbone emission and/or have "wavy" backbone. The characteristic
sizes of the inhomogeneities, more likely streamers, responsible
for the waving in our case laid between 0.1 and 0.5~R$_{\odot}$.

The discrepancy between visible CME velocity and retrieved from
the drift rate type II source radial velocity is explained by: 1 -
different velocities of the CME forehead and its flank where the
source is supposed to be located and 2 - non-radial propagation of
the source.

The statistical analysis of the HB sub-bursts parameters showed
that forward HB bursts had slower drift rates (-1.23~MHz~s$^{-1}$
in average) and were slightly longer (2.27~s) than the reverse
bursts (+1.8~MHz~s$^{-1}$ and 1.85~s respectively). In addition
the dependence of the drift rates on frequency for forward HB
sub-bursts found to be close to well known Alvarez and Haddock
(\citeyear{Alvarez73}) dependence for type III bursts while the
drift rates of reverse HB sub-bursts were proportional to the
frequency.

The HB structure expectedly appeared to be strongly polarized.
Moreover both forward and reverse HB sub-bursts were left-handed
polarized. Unlike the preceding type IIIs, whose sense of
polarization was opposite.

%

%
\begin{acks}
The work was partially performed under the support of the European
FP-7 project SOLSPANET (FP7-People-2010-IRSES-269299). The work of
M.P. was supported by the Austrian Fonds zur Förderung der
wissenschaftlichen Forschung (project P23762-N16).
 Conflict of
Interest: The authors declare that they have no conflict of
interest.
\end{acks}

%
%
\bibliographystyle{spr-mp-sola}

\begin{thebibliography}{29}
\ifx \bisbn   \undefined \def \bisbn  #1{ISBN #1}\fi \ifx \binits
\undefined \def \binits#1{#1}\fi \ifx \bauthor  \undefined \def
\bauthor#1{#1}\fi \ifx \batitle  \undefined \def \batitle#1{#1}\fi
\ifx \bjtitle  \undefined \def \bjtitle#1{\textit{#1}}\fi \ifx
\bvolume  \undefined \def \bvolume#1{\textbf{#1}}\fi \ifx \byear
\undefined \def \byear#1{#1}\fi \ifx \bissue  \undefined \def
\bissue#1{#1}\fi \ifx \bfpage  \undefined \def \bfpage#1{#1}\fi
\ifx \blpage  \undefined \def \blpage #1{#1}\fi \ifx \burl
\undefined \def \burl#1{\textsf{#1}}\fi \ifx \href  \undefined
\def \href#1#2{\textsf{#2}}\fi \ifx \doiurl  \undefined \def
  \doiurl#1{\href{http://dx.doi.org/#1}{\textsf{#1}}}\fi
\ifx \betal  \undefined \def \betal{\textit{et al.}}\fi \ifx
\binstitute  \undefined \def \binstitute#1{#1}\fi \ifx \bctitle
\undefined \def \bctitle#1{#1}\fi \ifx \beditor  \undefined \def
\beditor#1{#1}\fi \ifx \bpublisher  \undefined \def
\bpublisher#1{#1}\fi \ifx \bbtitle  \undefined \def
\bbtitle#1{\textit{#1}}\fi \ifx \bedition  \undefined \def
\bedition#1{#1}\fi \ifx \bseriesno  \undefined \def
\bseriesno#1{\textbf{#1}}\fi \ifx \blocation  \undefined \def
\blocation#1{#1}\fi \ifx \bsertitle  \undefined \def
\bsertitle#1{\textit{#1}}\fi \ifx \bsnm \undefined \def
\bsnm#1{#1}\fi \ifx \bsuffix \undefined \def \bsuffix#1{#1}\fi
\ifx \bparticle \undefined \def \bparticle#1{#1}\fi \ifx \barticle
\undefined \def \barticle#1{}\fi \ifx \botherref \undefined \def
\botherref#1{}\fi \ifx \url \undefined \def \url#1{\textsf{#1}}\fi
\ifx \bchapter \undefined \def \bchapter#1{}\fi \ifx \bbook
\undefined \def \bbook#1{}\fi \ifx \bcomment \undefined \def
\bcomment#1{#1}\fi \ifx \oauthor \undefined \def \oauthor#1{#1}\fi
\ifx \citeauthoryear \undefined \def \citeauthoryear#1{#1}\fi
\def \endbibitem {}
\ifx \bconflocation  \undefined \def \bconflocation#1{#1} \fi

\bibitem[\protect\citeauthoryear{{Alvarez} and {Haddock}}{1973}]{Alvarez73}
\begin{barticle}
\bauthor{\bsnm{{Alvarez}}, \binits{H.}},
\bauthor{\bsnm{{Haddock}}, \binits{F.T.}}: \byear{1973},
\batitle{{Solar Wind Density Model from km-Wave Type III Bursts}}.
\bjtitle{\solphys} \bvolume{29}, \bfpage{197}\,--\,\blpage{209}.
doi:\doiurl{10.1007/BF00153449}.
\end{barticle}
\endbibitem

\bibitem[\protect\citeauthoryear{{Braude} \textit{et~al.}}{1978}]{Braude78}
\begin{barticle}
\bauthor{\bsnm{{Braude}}, \binits{S.I.}}, \bauthor{\bsnm{{Megn}},
\binits{A.V.}}, \bauthor{\bsnm{{Riabov}}, \binits{B.P.}},
\bauthor{\bsnm{{Sharykin}}, \binits{N.K.}},
\bauthor{\bsnm{{Zhuk}}, \binits{I.N.}}: \byear{1978},
\batitle{{Decametric survey of discrete sources in the Northern
sky. I - The
  UTR-2 radio telescope: Experimental techniques and data processing}}.
\bjtitle{\apss} \bvolume{54}, \bfpage{3}\,--\,\blpage{36}.
doi:\doiurl{10.1007/BF00637902}.
\end{barticle}
\endbibitem

\bibitem[\protect\citeauthoryear{{Brazhenko}
  \textit{et~al.}}{2005}]{Brazhenko05}
\begin{barticle}
\bauthor{\bsnm{{Brazhenko}}, \binits{A.I.}},
\bauthor{\bsnm{{Bulatsen}}, \binits{V.G.}},
\bauthor{\bsnm{{Vashchishin}}, \binits{R.V.}},
\bauthor{\bsnm{{Frantsuzenko}}, \binits{A.V.}},
\bauthor{\bsnm{{Konovalenko}}, \binits{A.A.}},
\bauthor{\bsnm{{Falkovich}}, \binits{I.S.}},
\bauthor{\bsnm{{Abranin}}, \binits{E.P.}},
\bauthor{\bsnm{{Ulyanov}}, \binits{O.M.}},
\bauthor{\bsnm{{Zakharenko}}, \binits{V.V.}},
\bauthor{\bsnm{{Lecacheux}}, \binits{A.}},
\bauthor{\bsnm{{Rucker}}, \binits{H.}}: \byear{2005},
\batitle{{New decameter radiopolarimeter URAN-2}}.
\bjtitle{Kinematika i Fizika Nebesnykh Tel Supplement}
\bvolume{5}, \bfpage{43}\,--\,\blpage{46}.
\end{barticle}
\endbibitem

\bibitem[\protect\citeauthoryear{{Briand} \textit{et~al.}}{2008}]{Briand08}
\begin{barticle}
\bauthor{\bsnm{{Briand}}, \binits{C.}},
\bauthor{\bsnm{{Zaslavsky}}, \binits{A.}},
\bauthor{\bsnm{{Maksimovic}}, \binits{M.}},
\bauthor{\bsnm{{Zarka}}, \binits{P.}},
\bauthor{\bsnm{{Lecacheux}}, \binits{A.}},
\bauthor{\bsnm{{Rucker}}, \binits{H.O.}},
\bauthor{\bsnm{{Konovalenko}}, \binits{A.A.}},
\bauthor{\bsnm{{Abranin}}, \binits{E.P.}},
\bauthor{\bsnm{{Dorovsky}}, \binits{V.V.}},
\bauthor{\bsnm{{Stanislavsky}}, \binits{A.A.}},
\bauthor{\bsnm{{Melnik}}, \binits{V.N.}}: \byear{2008},
\batitle{{Faint solar radio structures from decametric
observations}}. \bjtitle{\aap} \bvolume{490},
\bfpage{339}\,--\,\blpage{344}.
doi:\doiurl{10.1051/0004-6361:200809842}.
\end{barticle}
\endbibitem

\bibitem[\protect\citeauthoryear{{Cairns} and {Robinson}}{1987}]{Cairns87}
\begin{barticle}
\bauthor{\bsnm{{Cairns}}, \binits{I.H.}},
\bauthor{\bsnm{{Robinson}}, \binits{R.D.}}: \byear{1987},
\batitle{{Herringbone bursts associated with type II solar radio
emission}}. \bjtitle{\solphys} \bvolume{111},
\bfpage{365}\,--\,\blpage{383}. doi:\doiurl{10.1007/BF00148526}.
\end{barticle}
\endbibitem

\bibitem[\protect\citeauthoryear{{Cane} and {White}}{1989}]{Cane89}
\begin{barticle}
\bauthor{\bsnm{{Cane}}, \binits{H.V.}}, \bauthor{\bsnm{{White}},
\binits{S.M.}}: \byear{1989}, \batitle{{On the source conditions
for herringbone structure in type II solar
  radio bursts}}.
\bjtitle{\solphys} \bvolume{120}, \bfpage{137}\,--\,\blpage{144}.
doi:\doiurl{10.1007/BF00148539}.
\end{barticle}
\endbibitem

\bibitem[\protect\citeauthoryear{{Carley} \textit{et~al.}}{2013}]{Carley13}
\begin{barticle}
\bauthor{\bsnm{{Carley}}, \binits{E.P.}}, \bauthor{\bsnm{{Long}},
\binits{D.M.}}, \bauthor{\bsnm{{Byrne}}, \binits{J.P.}},
\bauthor{\bsnm{{Zucca}}, \binits{P.}},
\bauthor{\bsnm{{Bloomfield}}, \binits{D.S.}},
\bauthor{\bsnm{{McCauley}}, \binits{J.}},
\bauthor{\bsnm{{Gallagher}}, \binits{P.T.}}: \byear{2013},
\batitle{{Quasiperiodic acceleration of electrons by a
plasmoid-driven shock in
  the solar atmosphere}}.
\bjtitle{Nature Physics} \bvolume{9},
\bfpage{811}\,--\,\blpage{816}. doi:\doiurl{10.1038/nphys2767}.
\end{barticle}
\endbibitem

\bibitem[\protect\citeauthoryear{{Dorovskyy}
  \textit{et~al.}}{2006}]{Dorovskyy06}
\begin{bchapter}
\bauthor{\bsnm{{Dorovskyy}}, \binits{V.V.}},
\bauthor{\bsnm{{Mel'Nik}}, \binits{V.N.}},
\bauthor{\bsnm{{Konovalenko}}, \binits{A.A.}},
\bauthor{\bsnm{{Rucker}}, \binits{H.O.}},
\bauthor{\bsnm{{Abranin}}, \binits{E.P.}},
\bauthor{\bsnm{{Lecacheux}}, \binits{A.}}: \byear{2006},
\bctitle{{Observations of Solar S-bursts at the decameter
wavelengths}}. In: \beditor{\bsnm{{Rucker}}, \binits{H.O.}},
\beditor{\bsnm{{Kurth}}, \binits{W.}}, \beditor{\bsnm{{Mann}},
\binits{G.}} (eds.) \bbtitle{Planetary Radio Emissions VI},
\bfpage{383}.
\end{bchapter}
\endbibitem

\bibitem[\protect\citeauthoryear{{Feng} \textit{et~al.}}{2012}]{Feng12}
\begin{barticle}
\bauthor{\bsnm{{Feng}}, \binits{S.W.}}, \bauthor{\bsnm{{Chen}},
\binits{Y.}}, \bauthor{\bsnm{{Kong}}, \binits{X.L.}},
\bauthor{\bsnm{{Li}}, \binits{G.}}, \bauthor{\bsnm{{Song}},
\binits{H.Q.}}, \bauthor{\bsnm{{Feng}}, \binits{X.S.}},
\bauthor{\bsnm{{Liu}}, \binits{Y.}}: \byear{2012}, \batitle{{Radio
Signatures of Coronal-mass-ejection-Streamer Interaction and
  Source Diagnostics of Type II Radio Burst}}.
\bjtitle{\apj} \bvolume{753}, \bfpage{21}.
doi:\doiurl{10.1088/0004-637X/753/1/21}.
\end{barticle}
\endbibitem

\bibitem[\protect\citeauthoryear{{Gopalswamy}}{2009}]{Gopalswamy09}
\begin{barticle}
\bauthor{\bsnm{{Gopalswamy}}, \binits{N.}}: \byear{2009},
\batitle{{Halo coronal mass ejections and geomagnetic storms}}.
\bjtitle{Earth, Planets, and Space} \bvolume{61},
\bfpage{595}\,--\,\blpage{597}.
\end{barticle}
\endbibitem

\bibitem[\protect\citeauthoryear{{Gopalswamy}}{2011}]{Gopalswamy11}
\begin{bchapter}
\bauthor{\bsnm{{Gopalswamy}}, \binits{N.}}: \byear{2011},
\bctitle{{Coronal mass ejections and their heliospheric
consequences}}. In: \bbtitle{Astronomical Society of India
Conference Series}, \bsertitle{Astronomical Society of India
Conference Series} \bseriesno{2}, \bfpage{241}\,--\,\blpage{258}.
\end{bchapter}
\endbibitem

\bibitem[\protect\citeauthoryear{{Gopalswamy}
  \textit{et~al.}}{2005}]{Gopalswamy05}
\begin{barticle}
\bauthor{\bsnm{{Gopalswamy}}, \binits{N.}},
\bauthor{\bsnm{{Aguilar-Rodriguez}}, \binits{E.}},
\bauthor{\bsnm{{Yashiro}}, \binits{S.}}, \bauthor{\bsnm{{Nunes}},
\binits{S.}}, \bauthor{\bsnm{{Kaiser}}, \binits{M.L.}},
\bauthor{\bsnm{{Howard}}, \binits{R.A.}}: \byear{2005},
\batitle{{Type II radio bursts and energetic solar eruptions}}.
\bjtitle{Journal of Geophysical Research (Space Physics)}
\bvolume{110}, \bfpage{12}. doi:\doiurl{10.1029/2005JA011158}.
\end{barticle}
\endbibitem

\bibitem[\protect\citeauthoryear{{Holman} and {Pesses}}{1983}]{Holman83}
\begin{barticle}
\bauthor{\bsnm{{Holman}}, \binits{G.D.}},
\bauthor{\bsnm{{Pesses}}, \binits{M.E.}}: \byear{1983},
\batitle{{Solar type II radio emission and the shock drift
acceleration of
  electrons}}.
\bjtitle{\apj} \bvolume{267}, \bfpage{837}\,--\,\blpage{843}.
doi:\doiurl{10.1086/160918}.
\end{barticle}
\endbibitem

\bibitem[\protect\citeauthoryear{{Kim} \textit{et~al.}}{2010}]{Kim10}
\begin{barticle}
\bauthor{\bsnm{{Kim}}, \binits{R.-S.}}, \bauthor{\bsnm{{Cho}},
\binits{K.-S.}}, \bauthor{\bsnm{{Moon}}, \binits{Y.-J.}},
\bauthor{\bsnm{{Dryer}}, \binits{M.}}, \bauthor{\bsnm{{Lee}},
\binits{J.}}, \bauthor{\bsnm{{Yi}}, \binits{Y.}},
\bauthor{\bsnm{{Kim}}, \binits{K.-H.}}, \bauthor{\bsnm{{Wang}},
\binits{H.}}, \bauthor{\bsnm{{Park}}, \binits{Y.-D.}},
\bauthor{\bsnm{{Kim}}, \binits{Y.H.}}: \byear{2010}, \batitle{{An
empirical model for prediction of geomagnetic storms using
  initially observed CME parameters at the Sun}}.
\bjtitle{Journal of Geophysical Research (Space Physics)}
\bvolume{115}, \bfpage{12108}. doi:\doiurl{10.1029/2010JA015322}.
\end{barticle}
\endbibitem

\bibitem[\protect\citeauthoryear{{Kong} \textit{et~al.}}{2012}]{Kong12}
\begin{barticle}
\bauthor{\bsnm{{Kong}}, \binits{X.L.}}, \bauthor{\bsnm{{Chen}},
\binits{Y.}}, \bauthor{\bsnm{{Li}}, \binits{G.}},
\bauthor{\bsnm{{Feng}}, \binits{S.W.}}, \bauthor{\bsnm{{Song}},
\binits{H.Q.}}, \bauthor{\bsnm{{Guo}}, \binits{F.}},
\bauthor{\bsnm{{Jiao}}, \binits{F.R.}}: \byear{2012}, \batitle{{A
Broken Solar Type II Radio Burst Induced by a Coronal Shock
  Propagating across the Streamer Boundary}}.
\bjtitle{\apj} \bvolume{750}, \bfpage{158}.
doi:\doiurl{10.1088/0004-637X/750/2/158}.
\end{barticle}
\endbibitem

\bibitem[\protect\citeauthoryear{{Magdaleni{\'c}}
  \textit{et~al.}}{2012}]{Jasmina12}
\begin{barticle}
\bauthor{\bsnm{{Magdaleni{\'c}}}, \binits{J.}},
\bauthor{\bsnm{{Marqu{\'e}}}, \binits{C.}},
\bauthor{\bsnm{{Zhukov}}, \binits{A.N.}}, \bauthor{\bsnm{{Vr{\v
s}nak}}, \binits{B.}}, \bauthor{\bsnm{{Veronig}}, \binits{A.}}:
\byear{2012}, \batitle{{Flare-generated Type II Burst without
Associated Coronal Mass
  Ejection}}.
\bjtitle{\apj} \bvolume{746}, \bfpage{152}.
doi:\doiurl{10.1088/0004-637X/746/2/152}.
\end{barticle}
\endbibitem

\bibitem[\protect\citeauthoryear{{Mann} and {Klassen}}{2005}]{Mann05}
\begin{barticle}
\bauthor{\bsnm{{Mann}}, \binits{G.}}, \bauthor{\bsnm{{Klassen}},
\binits{A.}}: \byear{2005}, \batitle{{Electron beams generated by
shock waves in the solar corona}}. \bjtitle{\aap} \bvolume{441},
\bfpage{319}\,--\,\blpage{326}.
doi:\doiurl{10.1051/0004-6361:20034396}.
\end{barticle}
\endbibitem

\bibitem[\protect\citeauthoryear{{McConnell}}{1982}]{McConnell82}
\begin{barticle}
\bauthor{\bsnm{{McConnell}}, \binits{D.}}: \byear{1982},
\batitle{{Spectral characteristics of solar S bursts}}.
\bjtitle{\solphys} \bvolume{78}, \bfpage{253}\,--\,\blpage{269}.
doi:\doiurl{10.1007/BF00151608}.
\end{barticle}
\endbibitem

\bibitem[\protect\citeauthoryear{{Melnik} \textit{et~al.}}{2004}]{Melnik04}
\begin{barticle}
\bauthor{\bsnm{{Melnik}}, \binits{V.N.}},
\bauthor{\bsnm{{Konovalenko}}, \binits{A.A.}},
\bauthor{\bsnm{{Rucker}}, \binits{H.O.}},
\bauthor{\bsnm{{Stanislavsky}}, \binits{A.A.}},
\bauthor{\bsnm{{Abranin}}, \binits{E.P.}},
\bauthor{\bsnm{{Lecacheux}}, \binits{A.}}, \bauthor{\bsnm{{Mann}},
\binits{G.}}, \bauthor{\bsnm{{Warmuth}}, \binits{A.}},
\bauthor{\bsnm{{Zaitsev}}, \binits{V.V.}},
\bauthor{\bsnm{{Boudjada}}, \binits{M.Y.}},
\bauthor{\bsnm{{Dorovskii}}, \binits{V.V.}},
\bauthor{\bsnm{{Zaharenko}}, \binits{V.V.}},
\bauthor{\bsnm{{Lisachenko}}, \binits{V.N.}},
\bauthor{\bsnm{{Rosolen}}, \binits{C.}}: \byear{2004},
\batitle{{Observations of Solar Type II bursts at frequencies
10-30 MHz}}. \bjtitle{\solphys} \bvolume{222},
\bfpage{151}\,--\,\blpage{166}.
doi:\doiurl{10.1023/B:SOLA.0000036854.66380.a4}.
\end{barticle}
\endbibitem

\bibitem[\protect\citeauthoryear{{Melnik} \textit{et~al.}}{2005}]{Melnik05}
\begin{barticle}
\bauthor{\bsnm{{Melnik}}, \binits{V.N.}},
\bauthor{\bsnm{{Konovalenko}}, \binits{A.A.}},
\bauthor{\bsnm{{Dorovskyy}}, \binits{V.V.}},
\bauthor{\bsnm{{Rucker}}, \binits{H.O.}},
\bauthor{\bsnm{{Abranin}}, \binits{E.P.}},
\bauthor{\bsnm{{Lisachenko}}, \binits{V.N.}},
\bauthor{\bsnm{{Lecacheux}}, \binits{A.}}: \byear{2005},
\batitle{{Solar Drift Pair Bursts in the Decameter Range}}.
\bjtitle{\solphys} \bvolume{231}, \bfpage{143}\,--\,\blpage{155}.
doi:\doiurl{10.1007/s11207-005-8272-4}.
\end{barticle}
\endbibitem

\bibitem[\protect\citeauthoryear{{Michalek} \textit{et~al.}}{2006}]{Michalek06}
\begin{barticle}
\bauthor{\bsnm{{Michalek}}, \binits{G.}},
\bauthor{\bsnm{{Gopalswamy}}, \binits{N.}},
\bauthor{\bsnm{{Lara}}, \binits{A.}}, \bauthor{\bsnm{{Yashiro}},
\binits{S.}}: \byear{2006}, \batitle{{Properties and
geoeffectiveness of halo coronal mass ejections}}. \bjtitle{Space
Weather} \bvolume{4}, \bfpage{10003}.
doi:\doiurl{10.1029/2005SW000218}.
\end{barticle}
\endbibitem

\bibitem[\protect\citeauthoryear{{Miteva} and {Mann}}{2007}]{Miteva07}
\begin{barticle}
\bauthor{\bsnm{{Miteva}}, \binits{R.}}, \bauthor{\bsnm{{Mann}},
\binits{G.}}: \byear{2007}, \batitle{{The electron acceleration at
shock waves in the solar corona}}. \bjtitle{\aap} \bvolume{474},
\bfpage{617}\,--\,\blpage{625}.
doi:\doiurl{10.1051/0004-6361:20066856}.
\end{barticle}
\endbibitem

\bibitem[\protect\citeauthoryear{{Reiner} \textit{et~al.}}{2003}]{Reiner03}
\begin{barticle}
\bauthor{\bsnm{{Reiner}}, \binits{M.J.}},
\bauthor{\bsnm{{Vourlidas}}, \binits{A.}}, \bauthor{\bsnm{{Cyr}},
\binits{O.C.S.}}, \bauthor{\bsnm{{Burkepile}}, \binits{J.T.}},
\bauthor{\bsnm{{Howard}}, \binits{R.A.}},
\bauthor{\bsnm{{Kaiser}}, \binits{M.L.}},
\bauthor{\bsnm{{Prestage}}, \binits{N.P.}},
\bauthor{\bsnm{{Bougeret}}, \binits{J.-L.}}: \byear{2003},
\batitle{{Constraints on Coronal Mass Ejection Dynamics from
Simultaneous Radio
  and White-Light Observations}}.
\bjtitle{\apj} \bvolume{590}, \bfpage{533}\,--\,\blpage{546}.
doi:\doiurl{10.1086/374917}.
\end{barticle}
\endbibitem

\bibitem[\protect\citeauthoryear{{Roberts}}{1959}]{Roberts59}
\begin{barticle}
\bauthor{\bsnm{{Roberts}}, \binits{J.A.}}: \byear{1959},
\batitle{{Solar Radio Bursts of Spectral Type II}}.
\bjtitle{Australian Journal of Physics} \bvolume{12},
\bfpage{327}. doi:\doiurl{10.1071/PH590327}.
\end{barticle}
\endbibitem

\bibitem[\protect\citeauthoryear{{Suzuki}, {Stewart}, and
  {Magun}}{1980}]{Suzuki80}
\begin{bchapter}
\bauthor{\bsnm{{Suzuki}}, \binits{S.}}, \bauthor{\bsnm{{Stewart}},
\binits{R.T.}}, \bauthor{\bsnm{{Magun}}, \binits{A.}}:
\byear{1980}, \bctitle{{Polarization of herringbone structure in
Type II bursts}}. In: \beditor{\bsnm{{Kundu}}, \binits{M.R.}},
\beditor{\bsnm{{Gergely}}, \binits{T.E.}} (eds.) \bbtitle{Radio
Physics of the Sun}, \bsertitle{IAU Symposium} \bseriesno{86},
\bfpage{241}\,--\,\blpage{245}.
\end{bchapter}
\endbibitem

\bibitem[\protect\citeauthoryear{{Vr{\v s}nak} and {Cliver}}{2008}]{Vrsnak08}
\begin{barticle}
\bauthor{\bsnm{{Vr{\v s}nak}}, \binits{B.}},
\bauthor{\bsnm{{Cliver}}, \binits{E.W.}}: \byear{2008},
\batitle{{Origin of Coronal Shock Waves. Invited Review}}.
\bjtitle{\solphys} \bvolume{253}, \bfpage{215}\,--\,\blpage{235}.
doi:\doiurl{10.1007/s11207-008-9241-5}.
\end{barticle}
\endbibitem

\bibitem[\protect\citeauthoryear{{Yashiro} \textit{et~al.}}{2008}]{Yashiro08}
\begin{barticle}
\bauthor{\bsnm{{Yashiro}}, \binits{S.}},
\bauthor{\bsnm{{Michalek}}, \binits{G.}},
\bauthor{\bsnm{{Akiyama}}, \binits{S.}},
\bauthor{\bsnm{{Gopalswamy}}, \binits{N.}},
\bauthor{\bsnm{{Howard}}, \binits{R.A.}}: \byear{2008},
\batitle{{Spatial Relationship between Solar Flares and Coronal
Mass
  Ejections}}.
\bjtitle{\apj} \bvolume{673}, \bfpage{1174}\,--\,\blpage{1180}.
doi:\doiurl{10.1086/524927}.
\end{barticle}
\endbibitem

\bibitem[\protect\citeauthoryear{{Zajtsev} \textit{et~al.}}{1998}]{Zaitsev98}
\begin{barticle}
\bauthor{\bsnm{{Zajtsev}}, \binits{V.V.}},
\bauthor{\bsnm{{Zlotnik}}, \binits{E.Y.}}, \bauthor{\bsnm{{Mann}},
\binits{G.}}, \bauthor{\bsnm{{Aurass}}, \binits{H.}},
\bauthor{\bsnm{{Klassen}}, \binits{A.}}: \byear{1998},
\batitle{{Efficiency of electron acceleration by shock waves in
the solar
  corona according to observational data on the fine structure of type II radio
  bursts.}}
\bjtitle{Radiophysics and Quantum Electronics} \bvolume{41},
\bfpage{107}\,--\,\blpage{114}. doi:\doiurl{10.1007/BF02679627}.
\end{barticle}
\endbibitem

\bibitem[\protect\citeauthoryear{{Zucca} \textit{et~al.}}{2012}]{Zucca12}
\begin{barticle}
\bauthor{\bsnm{{Zucca}}, \binits{P.}}, \bauthor{\bsnm{{Carley}},
\binits{E.P.}}, \bauthor{\bsnm{{McCauley}}, \binits{J.}},
\bauthor{\bsnm{{Gallagher}}, \binits{P.T.}},
\bauthor{\bsnm{{Monstein}}, \binits{C.}},
\bauthor{\bsnm{{McAteer}}, \binits{R.T.J.}}: \byear{2012},
\batitle{{Observations of Low Frequency Solar Radio Bursts from
the Rosse  Solar-Terrestrial Observatory}}. \bjtitle{\solphys}
\bvolume{280}, \bfpage{591}\,--\,\blpage{602}.
doi:\doiurl{10.1007/s11207-012-9992-x}.
\end{barticle}
\endbibitem

\end{thebibliography}

\newpage

\end{article}

\end{document}